\documentclass[12pt, preprint]{aastex}
\usepackage{amssymb}
\def\lapprox{$_<\atop{^\sim}$}
\def\grapprox{$_>\atop{^\sim}$}
\begin{document}
\title{H$_2$O and OH gas in the terrestrial planet-forming zones of protoplanetary disks}

\author{Colette Salyk\altaffilmark{1}}
\author{Klaus M. Pontoppidan\altaffilmark{1}}
\author{Geoffrey A. Blake\altaffilmark{1}}

\altaffiltext{1}{California Institute of Technology, Division of Geological and Planetary Sciences,
MS 150-21, Pasadena, CA 91125}

\author{Fred Lahuis\altaffilmark{2,3}}
\author{Ewine F. van Dishoeck\altaffilmark{3}}
\altaffiltext{2}{SRON National Institute for Space Research, P.O. Box 800, 9700 AV Groningen, The Netherlands}
\altaffiltext{3}{Leiden Observatory, Leiden University, P.O. Box 9513, 2300 RA Leiden, Netherlands}

\author{Neal J. Evans, II\altaffilmark{4}}
\altaffiltext{4}{Astronomy Department, University of Texas, TX 78712}

\begin{abstract}
We present detections of numerous 10\,--\,20\,$\mu$m H$_2$O emission lines from two protoplanetary
disks around the T Tauri stars AS 205A and DR Tau, obtained using the InfraRed Spectrograph on the Spitzer
Space Telescope.
Follow-up 3\,--\,5\,$\mu$m Keck-NIRSPEC data confirm the presence of abundant
water and spectrally resolve the lines. We also detect the P4.5 (2.934\,$\mu$m) and P9.5
(3.179\,$\mu$m) doublets of OH and $^{12}$CO/$^{13}$CO $v$=1$\rightarrow$0 emission in both sources.
Line shapes and LTE models suggest that the emission from all three molecules originates
between $\sim$0.5 and 5 AU,
and so will provide a new window for understanding
the chemical environment during terrestrial planet formation. LTE models also
imply significant columns of H$_2$O and OH in the inner disk atmospheres, suggesting physical transport of
volatile ices either vertically or radially; while the significant radial extent of the emission
stresses the importance of a more complete understanding of non-thermal excitation processes.
\end{abstract}

\keywords{astrochemistry --- circumstellar matter --- planetary systems: protoplanetary disks --- ISM: molecules}

\section{Introduction}

One of the most intriguing questions in the study of the formation of planets, and of terrestrial planets
in particular, is how water is transported to their surfaces, and whether or not water is a common ingredient
during their formation and early evolution. Spatially and spectrally resolved observations of water in extrasolar planetary systems and
protoplanetary disks would be instrumental in resolving these questions.
Unfortunately, detecting water in extrasolar planetary systems remains difficult \citep{Ehrenreich07} and
water in protoplanetary disks has been elusive, with only a few definitive ice \citep{Malfait98, Terada07}
and vapor \citep{Carr04} detections. Recently, however, the Spitzer
InfraRed Spectrograph (IRS) has begun to reveal
warm water vapor emission lines from disks
around classical T Tauri stars (cTTs, Carr \& Najita 2008).

In this Letter, we present high-resolution spectroscopy of numerous water vapor emission lines from
protoplanetary disks around two cTTs: DR Tau and AS 205A.  Spectra were obtained
using both the Spitzer-IRS and NIRSPEC \citep{McLean98} at the Keck II telescope.
Accretion rates onto these stars are high, which causes the inner disks to be significantly hotter than
in more quiescent systems.  DR Tau is known to have variable mass accretion, with measured rates between
$0.3$ and $79\times 10^{-7}\,M_{\odot}\,\rm yr^{-1}$ \citep{Gullbring00,JohnsKrull02}. AS 205A has an accretion rate of $7.2\times 10^{-7}\,M_{\odot}\,\rm yr^{-1}$ and is the primary component of a 1.3\arcsec\ triple system, with the secondary being a spectroscopic binary \citep{Eisner05}.
However, it is likely that the physical separation of AS 205A and B is significantly larger than the
projected separation and that, therefore, AS 205B does not affect the (inner) disk of AS 205A \citep{Andrews07}.

\section{Observations}
The Spitzer IRS Short-High (SH) module spectrum of AS 205A was taken as part of the
``cores to disks'' (c2d) Legacy program \citep{Evans03,kessler06},
while the spectrum of DR Tau, observed as part of the guaranteed time observations \citep{Houck04},
was obtained from the Spitzer archive. The 2D basic calibrated data images were reduced
using the c2d pipeline as described in \citet{Lahuis06}.

To confirm the detections as well as measure the gas emission line profiles, we obtained
high resolution ($\lambda/\Delta\lambda$=$25000$) L-band spectra using NIRSPEC
in a single grating setting, of which two orders, centered at
2.92\,$\mu$m and 3.16\,$\mu$m, respectively, have identifiable H$_2$O and OH emission features.
We also collected CO $v$=1$\rightarrow$0 M-band spectra ($\sim$5\,$\mu$m) as part of an
on-going survey (c.f. \citealp{Blake04}, \citealp{Salyk07}).
The data were reduced in a standard way.  Division by the spectra of standard stars
(HR 1620 (A7V) for DR Tau; HR 5993 (B1V)
for AS 205A) was used to correct for telluric absorption, which can be
poor when the absorption is high. Thus, regions of the spectrum with $<$70\%
atmospheric transmission were removed. Fluxes were calibrated using the photometry and
spectral types of the standards, and are generally accurate to {\lapprox}20\%.

The NIRSPEC lines are narrow (FWHM$\lesssim$ 35 km s$^{-1}$), spatially unresolved,
and centered at the stellar $v_{\rm LSR}$ \citep{Herbig88, Thi01} to within our uncertainty (a few
km s$^{-1}$) --- all consistent with disk emission.  Because the disks are optically thick and heated from
above, the line emission likely arises from the upper disk atmosphere \citep{Najita03}.

\section{Results}

\subsection{Spitzer-IRS}\label{sec:spitzer}
\begin{figure}
 \plotone{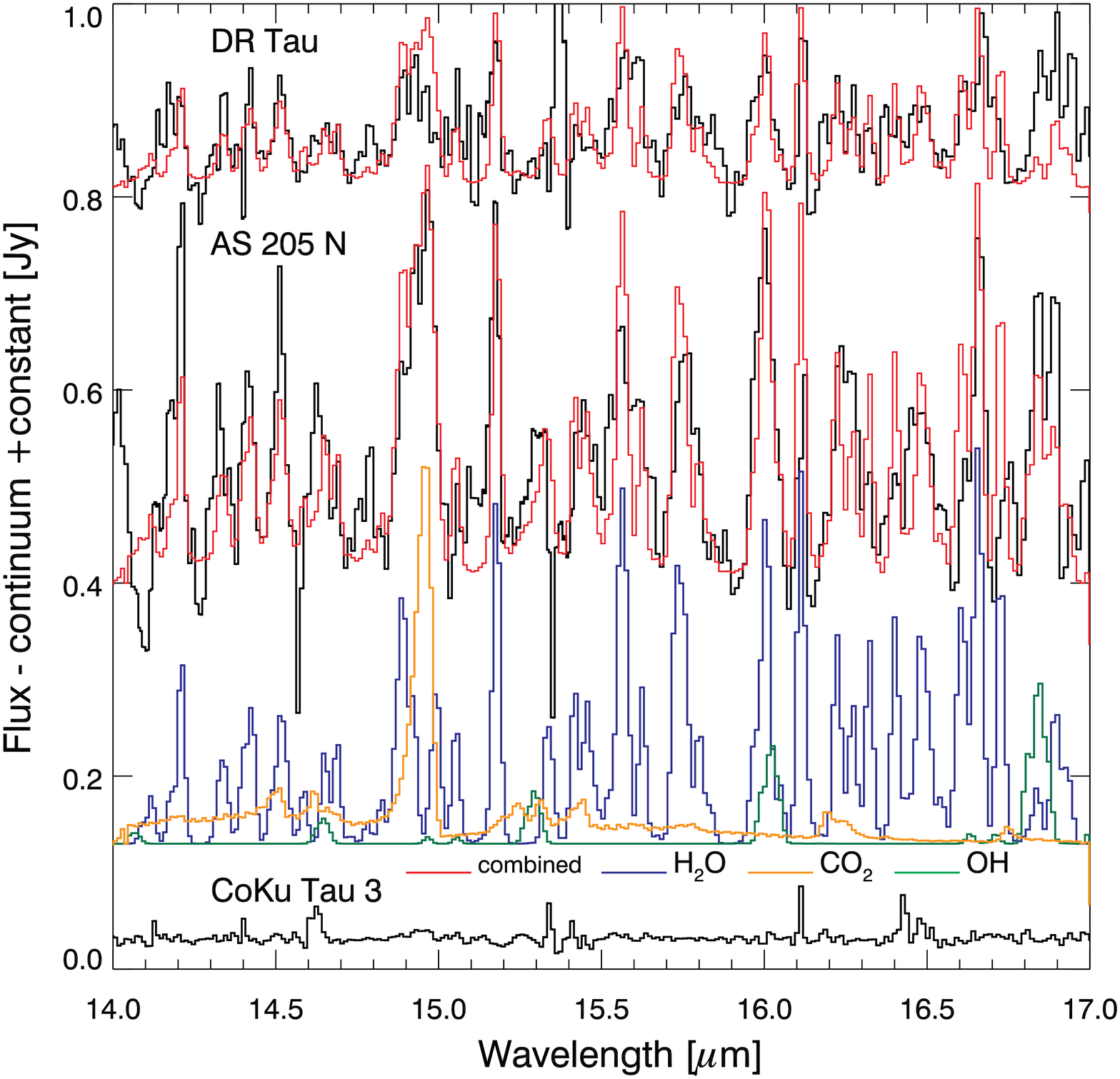}
  \caption{A portion of the Spitzer-IRS data (in black) and molecular emission models (in red).
  The three components of the AS 205A model are shown in color below.
  The largely featureless spectrum of CoKu Tau/3  is included to demonstrate the dynamic range achievable by the IRS and the level of systematics,
  such as residuals from the de-fringing process. For both spectra, splines have been used to remove
  broad continuum features from silicates. Line-to-continuum excesses are $\sim$\,5\,--\,10\%.}
  \label{IRS_H2O}
\end{figure}

The IRS spectra of AS 205A and DR Tau show a large number of water emission lines (see Figure \ref{IRS_H2O}),
the vast majority pure rotational transitions with quantum numbers in the
range 20\,--\,50.  The $R$=600 resolution of the SH module results in unresolved, and significantly blended, lines.
Consequently, the identification with water is achieved via a simple, isothermal LTE ring model
computed using the HITRAN 2004 database \citep{Rothman05}.  The ring model consists of a single
temperature ($T$) and column density ($N$) gas in Keplerian rotation. Model parameters include $T$,
$N$, solid angle ($\Omega$), stellar mass ($M_\star$), inclination ($i$) and characteristic radius ($r$).
In addition, the local line broadening is assumed to be Gaussian, with specified width ($\sigma$).
$M_\star$ and $i$ were fixed to values taken from the literature
\citep{Clarke00,Kitamura02,Eisner05,Andrews07}. The spectrally unresolved IRS data cannot completely constrain
the remaining model parameters, but do require high $T$ ($T\gtrsim 500$ K)
and $N$ ($N\gtrsim 10^{17}$ cm$^{-2}$ for $\sigma$$\sim$2 km s$^{-1}$).
Figure \ref{IRS_H2O} compares the IRS spectra with a model in which $T$ and $N$ are constrained using
the NIRSPEC data described below, but where $\Omega$ has been increased by a factor of 2.

The addition of CO$_2$ (and, for AS 205A, OH) improves our fits, as shown by the AS 205A
model in Fig. \ref{IRS_H2O}.  With $T$ and $\Omega$ fixed, we find H$_2$O:CO$_2$=40 for DR Tau and
20 for AS 205.  Detections of these and other molecules will be discussed in detail elsewhere.

\subsection{NIRSPEC}
\label{sec:nirspec}
\begin{figure}
  \plotone{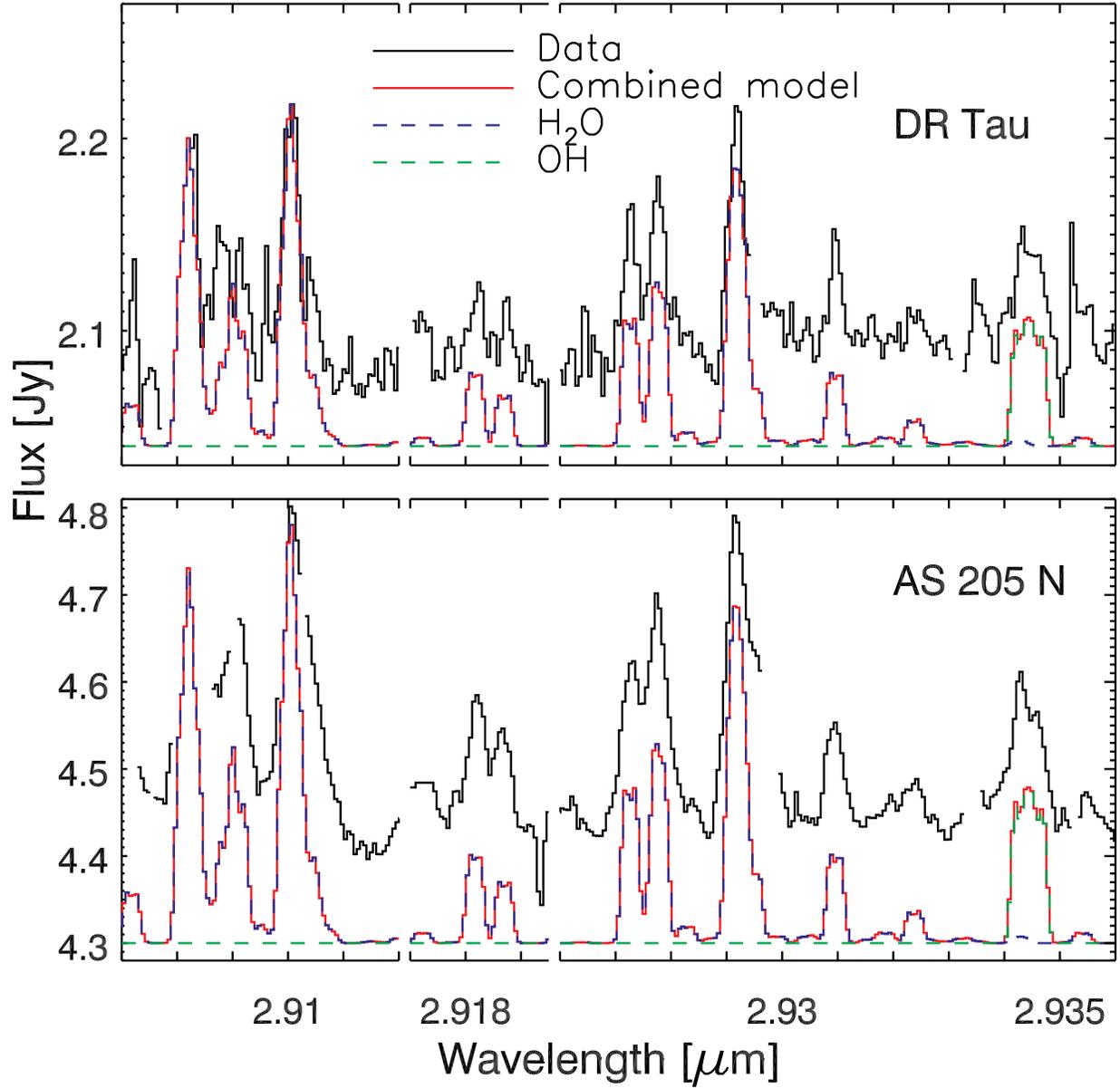}
  \caption{Comparison between portions of the NIRSPEC L-band data and best-fit H$_2$O/OH disk models (offset).
}
  \label{nirspec}
  \vskip 0.05in
\end{figure}

\begin{figure}
\plotone{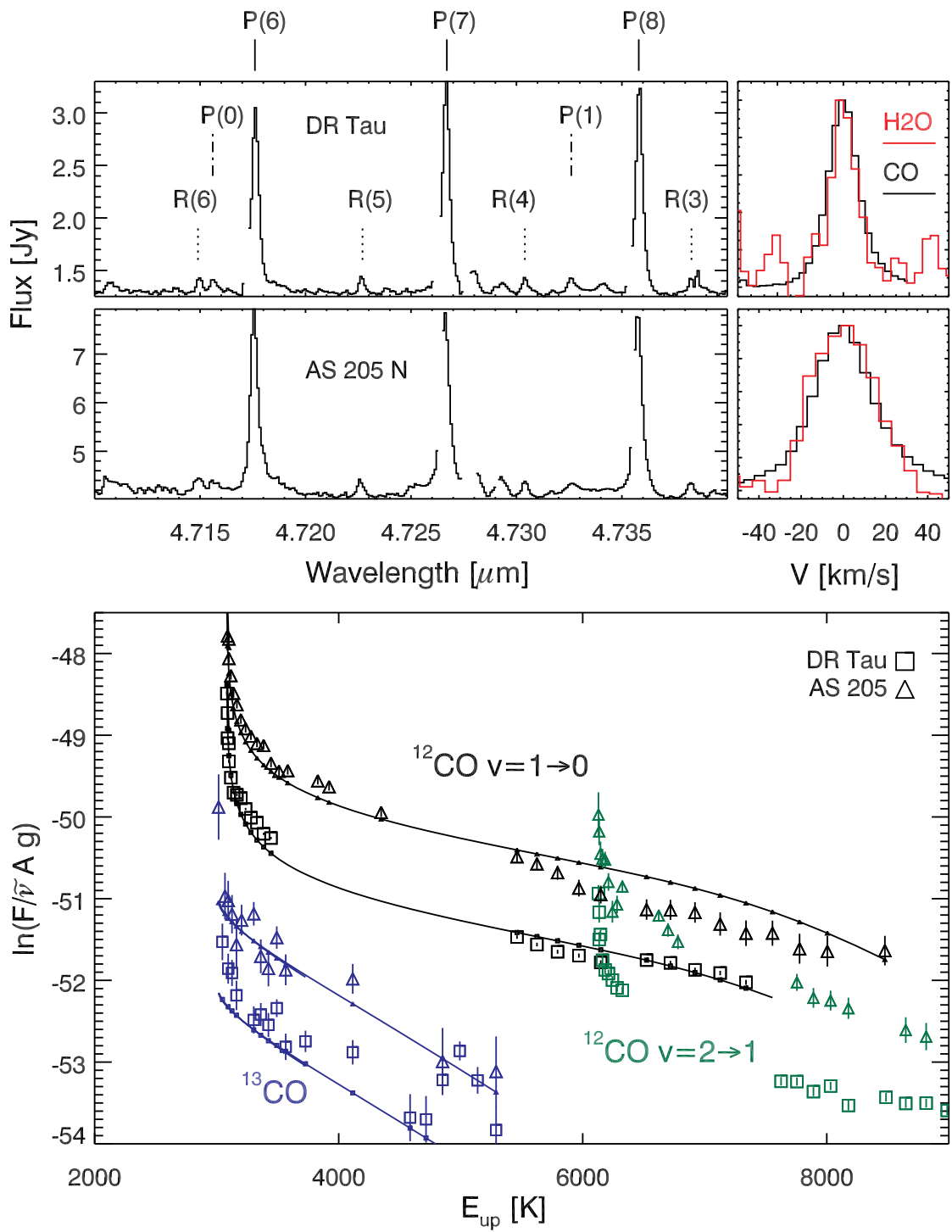}
  \caption{Top, left: a portion of the NIRSPEC M-band spectra.  $^{12}$CO/$^{13}$CO $v$=1$\rightarrow$0 transitions
  are marked with solid and dashed lines, respectively.  $^{12}$CO $v$=2$\rightarrow$1 transitions are marked
  with dot-dashed lines.  Top, right: CO and H$_2$O emission lines are overplotted.  Bottom: an excitation diagram with
  1$\sigma$ error bars and best-fit models to the $v$=1$\rightarrow$0 transitions. }
\label{rotfit}
\end{figure}

While tropospheric water vapor blocks significant portions of incoming radiation near 3\,$\mu$m,
there are a handful of rovibrational lines that are optically thin in the atmosphere but readily excited at
temperatures typical of the inner $\sim$\,1 AU of circumstellar disks.  This is  illustrated
in Figure \ref{nirspec}, in which spectral regions with $<$70\% atmospheric transmission have been
removed, yet many prominent (line-to-continuum\,$\sim$\,5\,--\,10\%) H$_2$O emission features remain.
These are primarily $\nu_3=1\rightarrow0$ lines, with upper level energies of $\sim$\,5,000\,--\,10,000 K.
A few weaker (line-to-continuum\,$\sim$\,1\%) features are nominally detected from 3.13\,--\,3.18 $\mu$m
and are used as model constraints.  We also detect the P4.5 (2.9344\,$\mu$m;
$T_\mathrm{up}$\,$\sim$\,$5400$ K) and P9.5 (3.1788\,$\mu$m; $T_\mathrm{up}$\,$\sim$$\,7500$ K) OH
doublets. The M-band spectra include portions of the $^{12}$CO/$^{13}$CO $v$=1$\rightarrow$0
and $^{12}$CO v=2$\rightarrow$1 P- and R-branches
(Figure \ref{rotfit}).

We fit all spectra with the ring models described in \S\ref{sec:spitzer}.
Because the CO lines span a large range of excitation energies,
across which they transition from optically thick to optically thin,
$N$, $T$, and $\Omega$ can be determined uniquely.  Also, a constraint on $N_\mathrm{CO}$ is provided
by the  $^{13}$CO lines (assuming a $^{12}$CO:$^{13}$CO ratio, here fixed at
77, c.f. \citealp{Blake04}).  Thus, we began by fitting the CO emission by comparison
with an excitation diagram of the $v$=1$\rightarrow$0 transitions (Fig. \ref{rotfit}), as described in \citet{Salyk07}.
For our nominal H$_2$O and OH fits,  we then fixed $\Omega$ and $T$ to the CO-derived values to
determine $N$ -- a reasonable, though not ideal, assumption, given the similarity in line shapes (see Fig. \ref{rotfit}).

Our nominal ring model is shown in Figure \ref{nirspec}, with best-fit parameters in Table \ref{fit_parameters},
and uncertainties of order $\Delta N$\,$\sim$\,$30\%$, $\Delta T$\,$\sim$\,$15\%$
and $\Delta\Omega$\,$\sim$\,$10\%$. We have fixed $r=3$ AU to roughly match the emission profiles since the
lineshape does not affect the integrated flux; we discuss lineshapes below. $\Omega$ corresponds to areas
of 0.2 and 0.4 AU$^2$ for DR Tau and AS 205, respectively, at distances of 140 and 120 pc.  Uncertainties
in the nominal H$_2$O and OH fits are linked to CO uncertainties, but if $T$ is instead allowed to vary,
the H$_2$O spectra are consistent with $T$$\sim$\,900\,--\,1200/900\,--\,1100 K for DR Tau/AS 205A and ranges in
$N$ of about 1 order of magnitude.  For OH, we can only constrain $T$ {\grapprox}900 K.

The model results depend crucially on the assumed local line width ($\sigma$), which is unknown.  For our nominal
model, we have adopted $\sigma=2\,$km s$^{-1}$ --- the sound speed for H$_2$ at 1000 K.  However, we have
also tested values between $\sim$0.3 and 10 km s$^{-1}$.  We find that the opacity needs
to remain similar from fit to fit, so that a change in $\sigma^2$ requires a proportional change in $N$.
To maintain the flux level, the total number of molecules must remain similar, so an increase in $\sigma$ implies a decrease in
$\Omega$.  Accordingly, the CO:H$_2$O ratio changes by no more than a factor of $\sim$ 2 between models.
 (If $T_{\mathrm{H_2O}}$ is not fixed, however, this ratio may change). In addition, $T$ is only weakly
affected by $\sigma$, since it is set by the overall shape of the spectrum, rather than by the absolute flux levels.

NIRSPEC spectrally resolves the molecular emission, allowing an investigation of the disk gas kinematics.
In Fig. \ref{rotfit} we compare the unblended H$_2$O transition at 2.931 $\mu$m with a CO line composite
constructed from an average of all $v$=1$\rightarrow$0 lines. In general, Keplerian disks should
produce double-peaked emission profiles, with peak separation set by the inner and outer emission radii
($r_\mathrm{in}$, $r_\mathrm{out}$), as well as the temperature gradient \citep{Horne86}.
The CO and H$_2$O line profiles, however, are more Gaussian, or even Lorentzian in shape,
with relatively wide wings and a narrow peak.  The narrow core could be produced by a ring with a very
large $r_\mathrm{in}$ ($>$10 AU) if the peak-to-peak separation is equal to the NIRSPEC resolution (given the $M_\star$
and $i$ in Table \ref{fit_parameters}). However, this is difficult to reconcile with the high $T$
derived from the model fits and the observed line wings.

Another way to produce a single peak would be to have the line emission extend to significant radii ($r_\mathrm{out}$ of
order 7 AU, for a temperature profile that is constant with radius), such that the contribution from
low-velocity portions of the disk infills the center of the line profile.  However, this model produces too much
flux to match observations.  Additionally,  passive disk models predict a steep temperature decline with radius.
A more promising solution may be that some flux is produced via fluorescence, in which molecules in the flared outer disk
(typically out to $\sim$5 AU) intercept and re-radiate inner disk IR continuum, thus filling in the
line profiles at low velocity \citep{Blake04}. With canonical disk flaring and temperature structure,
we find that a resonance fluorescence model with 100\% scattering efficiency can indeed produce
fluxes comparable to those of thermal emission; a more thorough investigation is left as future work.

In any case, the amount of flux in the line profile wings is consistent with emission arising in a
Keplerian disk with a small inner radius.  With this assumption, the line wing velocity can be used to estimate $r_\mathrm{in}$.
Because turbulent motion and disk structure can both have non-negligible
effects on the line shapes, we adopt here the velocity, $v_\mathrm{in}$, at 2.2 times the Half Width at Half
Maximum --- a value found to be appropriate for the moderate-inclination disk around GM Aur \citep{Salyk07}.
To compute $v_\mathrm{in}$ we use Gaussian fits to the CO and H$_2$O lines discussed above, as well as the average
of two-Gaussian fits to the OH emission doublets, deconvolved with the NIRSPEC instrument response function. If turbulent
velocities are low, radii associated with $v_\mathrm{in}$ will
be overestimates, and so we also compute $v$ at 3$\sigma$, which should represent a lower limit
to $r_\mathrm{in}$ (see Table \ref{fit_parameters}).

\section{Discussion}

We have convincingly detected H$_2$O emission in the 10\,--\,20 $\mu$m region with Spitzer-IRS as well as
H$_2$O and OH emission near 3 $\mu$m with NIRSPEC, arising from the disk atmospheres surrounding two cTTs.
By combining these data with constraints from spectrally resolved CO $v$=1$\rightarrow$0
emission, we find that the excitation temperatures are typical of terrestrial planet
forming regions ($\sim$\,1000 K).  Additionally, line wing velocities imply inner emission radii no
larger than 1 AU for all molecules.  Therefore, further observations and analyses of these
species may provide a new window into terrestrial planet forming regions.

Observations of H$_2$O in disks may also provide constraints on disk evolution and water transport.
In a disk in which water vapor transport is controlled by diffusion, the presence of a condensation
front (`snow line') can potentially dry out an inner disk in as little as $10^{5}$ yr, for typical disk viscosities
\citep{Stevenson88}.  The snow line for the early solar system is estimated to be near 3 AU, but
would be further out for these high accretion rate stars \citep{Lecar06}.   Therefore, the presence of significant
water vapor inside 1 AU may be evidence for inward radial migration, or upward mixing, of icy solids, followed by
evaporation.  With both solid migration and diffusion controlling water transport, water vapor concentrations become sensitive to an
array of parameters, including disk viscosity and planetesimal growth rates, and water concentrations may trace disk evolution \citep{Ciesla06}.

The detection of strong OH emission is also interesting, for OH is known to be an
important ingredient to the chemistry in the inner regions of protoplanetary disks. It also
acts as a coolant near the disk surface \citep[e.g.][]{Dullemond07}, and
photolysis of OH to O($^1$D) has been invoked to explain the observed strengths
of the 6300\,\AA~O I line \citep{Acke05}.  Detections of OH in these and other disks will be discussed
in greater detail in Mandell et al., in prep.

An understanding of the local line broadening in disks will be crucial for obtaining accurate absolute column
densities of molecular species, as smaller local line widths require lower column densities and vice versa.
Nevertheless, we find that molecular ratios are more robust, and remain similar when utilizing
different disk models with the nominal fits having CO:H$_2$O\,$\sim$\,10--15 and H$_2$O:OH\,$\sim$\,3--4,
provided $T_\mathrm{H_2O} = T_\mathrm{OH} = T_\mathrm{CO}$. Stronger constraints on both
$T$ and $N$ will require high line-to-continuum observations across a wider wavelength range, but the
available data reveal that H$_2$O:CO is similar to, or slightly higher than that in the
dense clouds out of which disks form \citep{Boogert04}.  Vertically integrated H$_2$O:OH ratios have been predicted
to be much higher ($>10^5$) near 1 AU \citep{Markwick02}, but significant OH abundances are expected
near the C$^+$/C/CO transition zone at large scale heights \citep{kamp04}.  Also, it is important to keep
in mind that disk transport is not included in current chemical models.


Our analysis of the L- and M-band spectra thus far has assumed thermal
excitation of the vibrationally excited states, but this need not be the
case.  $\Delta v$=1 emission can also be driven by the stellar UV output
or by the absorption of IR photons from the star or disk, followed by
fluorescence \citep{DelloRusso04, Blake04}. For OH, vibrationally excited
emission can also result from the (photo)dissociation of H$_2$O followed
by prompt emission \citep{Bonev06}. In fact, fluorescence may be necessary
to explain observed line profile shapes (see \S\ref{sec:nirspec}),
and could explain the surprisingly strong OH emission.  A better
understanding of these processes may therefore prove essential for
relating observed line fluxes to disk column densities.

If emission in the line wings truly represents emission from the
disk inner radius, relatively large inner emission radii are implied for CO, OH, and H$_2$O.   Both  DR Tau and
AS 205A have had their dust inner rim sizes measured with the Keck Interferometer.  At
$0.11\pm{0.03}$ and $0.07\pm{0.01}$ AU, respectively, they are consistent
with the location of dust sublimation \citep{Akeson05, Eisner05}, while the inner emission radii we derive are at
least 3 times larger.  Perhaps these high accretion-rate stars can stir up the
inner disk enough to erase the vertical temperature gradients necessary for formation of emission
lines.  Another possibility may be that replenishment rates cannot keep up with photo-dissociation at
the disk surface.

The ease with which H$_2$O and OH are now detected in the terrestrial planet-forming zones of certain 
disks holds the promise of more extensive and detailed studies.  Although the ground based echelle 
observations presented here probe only the inner disk atmospheres, future high spectral resolution 
follow-up studies of Spitzer-detected water emission from protoplanetary disks will provide constraints 
at a variety of disk radii and vertical depths. Additionally, spectro-astrometric observations are being 
used to directly constrain emitting location and structure of the molecular gas \citep{Pontoppidan08}.  
Eventually, a suite of constraints combined with non-LTE radiative transfer models will allow us to 
rigorously address such far-reaching questions as:  How does the water vapor abundance vary within and 
between disks?  What constraints can be put on models for the chemistry and transport of volatiles in 
disks, and what implications will these have for the early evolution of the Solar System?

$~~~$

\acknowledgements{
We are grateful to Joan Najita for discussions that prompted this work, and the anonymous
referee for a thoughtful review. Support
was provided in part by the Spitzer Space Telescope Legacy Science Program (NASA Contract
Numbers 1224608 and 1230779 issued by the Jet Propulsion Laboratory, California Institute of
Technology under NASA contract 1407). KMP is supported by NASA through Hubble Fellowship grant
\#01201.01 awarded by the Space Telescope Science Institute, which is operated by the Association
of Universities for Research in Astronomy, Inc., for NASA, under contract NAS 5-26555.
FL and EvD are supported by NOVA and NWO-Spinoza grants.  Some of the
results herein were obtained at the W.M. Keck Observatory, which is operated as a scientific
partnership among the California Institute of Technology, the University of California and NASA.
The Observatory was made possible by the generous financial support of the W.M. Keck Foundation.
}

\begin{table}
\centering
\caption{Fits to NIRSPEC data}
\begin{tabular}{lll}
\hline
\hline
&  DR Tau & AS 205A\\
\hline
CO $v_\mathrm{in}$, $v_{3\sigma}$\ \ [km\,s$^{-1}$]  &27, 32 & 35, 40  \\
CO $r_\mathrm{in}$, $r_{3\sigma}$\ \  [AU] &0.8, 0.6&0.5, 0.4  \\
H$_2$O $v_\mathrm{in}$, $v_{3\sigma}$ [km\,s$^{-1}$] & 24, 28 & 36, 42    \\
H$_2$O $r_\mathrm{in}$, $r_{3\sigma}$ [AU] &   1.0,   0.7  &0.4, 0.3\\
OH $v_\mathrm{in}$, $v_{3\sigma}$\ \  [km\,s$^{-1}$] & 28, 32&34, 39  \\
OH $r_\mathrm{in}$, $r_{3\sigma}$\  \, [AU] &0.7, 0.5&0.5,  0.4 \\
\hline
$r$[AU] \tablenotemark{a}  & 3 & 3 \tablenotemark{b}\\
$\sigma$  [km s$^{-1}$] \tablenotemark{a} & 2 & 2\\
$M_\star$ [$M_\odot$] \tablenotemark{a}& 0.76 & 1.2\\
$i [^\circ]$\tablenotemark{a} &67&47\\
$\Omega_\mathrm{CO, H_2O, OH}$ [sr] &  $2\times10^{-16}$         & $6\times10^{-16}$ \\
$T_\mathrm{CO, H_2O, OH}$ [K]               & 1000 & 1000 \\
$N_\mathrm{CO}$\ \ \  [cm$^{-2}$]    &$8\times 10^{18}$ & $9\times 10^{18}$ \\
$N_\mathrm{H_2O}$\,  [cm$^{-2}$]  &$8\times10^{17}$ &$6\times 10^{17}$ \\
$N_\mathrm{OH}$\ \ \   [cm$^{-2}$]  & $2\times 10^{17}$ & $2\times 10^{17}$\\
\hline
\tablenotetext{a}{Fixed}
\tablenotetext{b}{Chosen to approximately match line shapes}
\end{tabular}
\label{fit_parameters}
\vskip 0.05in
\end{table}

\end{document}